\documentclass[]{aa}  
\usepackage{natbib}
\bibpunct{(}{)}{;}{a}{}{,} 
\usepackage{graphicx}
\usepackage{caption}
\usepackage{subcaption}
\usepackage{txfonts}
\usepackage[dvipsnames]{xcolor}
\usepackage[colorlinks=true,citecolor=RoyalBlue]{hyperref}
\begin{document}

   \title{Composition of giant planets: the roles of pebbles and planetesimals}

   \author{C. Danti
          \inst{1,2}
          \and
          B. Bitsch\inst{2,3}
          \and
          J. Mah\inst{2}
          }

   \institute{Center for Star and Planet Formation, Globe Insitute, Øster Voldgade 5, 1350 Copenhagen, Denmark\\
              \email{danticlaudia@gmail.com}
         \and
             Max-Planck-Institut für Astronomie, Königstuhl 17, 69117 Heidelberg, Germany 
            \and
            Department of Physics, University College Cork, Cork, Ireland
             }

   \date{Received July 2023; accepted September 2023}

  \abstract
   {One of the current challenges of planet formation theory is to explain the enrichment of observed exoplanetary atmospheres.
   While past studies have focused on scenarios where either pebbles or planetesimals are the main drivers of heavy element enrichment, we combine here both approaches to understand whether the composition of a planet can constrain its formation pathway. We study three different formation scenarios: pebble accretion, pebble accretion with planetesimal formation inside the disc, combined pebble and planetesimal accretion.
   We use the \textsc{chemcomp} code to perform semi-analytical 1D simulations of protoplanetary discs, including viscous evolution, pebble drift, and simple chemistry to simulate the growth of planets from planetary embryos to gas giants as they migrate through the disc, while simultaneously tracking their composition.
   Our simulations confirm that the composition of the planetary atmosphere is dominated by the accretion of gas vapour enriched by inward drifting and evaporating pebbles. Including planetesimal formation hinders this enrichment, because many pebbles are locked into planetesimals and cannot evaporate and enrich the disc. This results in a dramatic drop of the accreted heavy elements both in the planetesimal formation and accretion case, demonstrating that planetesimal formation needs to be inefficient in order to explain planets with high heavy element content.
   On the other hand, accretion of planetesimals enhances the refractory component of the atmosphere, leading to low volatile to refractory ratios in the atmosphere, in contrast to the majority of pure pebble simulations. However, low volatile to refractory ratios can also be achieved in the pure pebble accretion scenario, if the planet migrates all the way into the inner disc and accretes gas that is enriched in evaporated refractories.
   Distinguishing these two scenarios requires knowledge about the planet's atmospheric C/H and O/H ratios, which are much higher in the pure pebble scenario compared to the planetesimal formation and accretion scenario. 
   This implies that a detailed knowledge of the composition of planetary atmospheres could help to distinguish between the different formation scenarios.}

   \keywords{Planets and satellites: composition -- 
             Planets and satellites: formation -- 
             Planets and satellites: gaseous planets -- 
             Protoplanetary disks 
            }

   \maketitle
%

\section{Introduction}

The exact mechanism of planet formation is still under debate, despite the number of confirmed exoplanets being now more than 5000 \footnote{https://exoplanetarchive.ipac.caltech.edu/, accessed 11 July 2023.}. The two models in the core accretion scenario are planet formation via planetesimal \citep{Pollack_1996} or via pebble accretion \citep{Ormel_2010, Lambrechts_2012}.
The planetesimal accretion scenario is based on the idea that the cores of the planets form by accretion of planetesimals in the size range of sub-kilometer to several tens of kilometres and then subsequently undergo runaway gas accretion. This scenario faces a main issue regarding cold gas giant formation: the planetesimal accretion rate drops significantly with the distance from the central star  due to long collisional time scales., resulting in a too low accretion rate to form a sufficiently big core during the disc gas phase that would allow runaway gas accretion \citep{Tanaka_1999, Levison_2010, Johansen2019}.

In the pebble accretion scenario that planetary growth is driven by the accretion of mm to cm sized pebbles \citep{Ormel_2010, Johansen_2010, Lambrechts_2012}. This process is much faster, resulting in planetary growth rates that are several orders of magnitude larger than planetesimal accretion rates, allowing more efficient gas giant formation.
The pebble accretion mechanism can also be efficient in the outer disc \citep[e.g.][]{Lambrechts_2012, Lambrechts_2014, Bitsch_2015}, while planetesimal accretion is rather inefficient at these large distances \citep[e.g.][]{Pollack_1996, Tanaka_1999, Johansen2019, Emsenhuber2020}.

In the past these models have been constrained via observations of planetary masses, radii and their orbital distances. However, these planet formation models are now challenged by a new component: measurements of atmospheric abundances \citep[e.g.][]{Line_2021, Pelletier_2021, Bean_2023}. Especially the data from the James Webb Space Telescope (JWST) will push this field forward, with first interesting results already coming in \citep[e.g.][]{Bean_2023}.
It is thought that the atmospheric composition of planets holds the key to their formation location, with particular importance placed on the C/H, O/H and C/O ratios of the atmospheres, because they vary with orbital distance from the star due to the evaporation of different oxygen and carbon bearing species like H$_2$O, CO$_2$, CH$_4$ and CO \citep[e.g.][]{oeberg2011, Madhusudhan2017, Booth_2019,Bitsch2021_1, Molliere_2022}.

In addition to atmospheric abundances, the bulk abundances of the planet have also gained attention as a potential metric to constrain planet formation models \citep{Thorngren2016}, where planetesimal-driven scenarios seem to have trouble explaining the large heavy element contents \citep[e.g.][]{Venturini_2020}, while pebble-based scenarios seem to be more promising \citep{Bitsch2021_1, Morbidelli_2023}.

Previous planet formation models usually assumed either that all material is available in form of pebbles \citep[e.g.][]{Lambrechts_2012, Bitsch_2015, Bitsch2021_1} or completely in form of planetesimals \citep[e.g.][]{Mordasini_2012, Emsenhuber2020}. These approaches ignore either that a planetary embryo has to first form starting from pebbles or that the planetesimal formation is not $100\%$ efficient \citep[e.g.][]{Johansens_2014}. 

We present here a model that includes planetesimal formation from an inward flux of pebbles following the recipe in \citep{Lenz2019} presented in Appendix \ref{app:pla_form} with the goal to simulate the composition of growing giant planets. In particular, we analyse three possible formation scenarios: planetary growth via pebble and subsequent gas accretion, growth by pebble and gas accretion with the possibility of forming planetesimals in the disc but not accreting them, and finally a combined growth scenario via pebble and planetesimal accretion.

We do not take a pure planetesimal scenario into account, because planetesimals form from inward drifting pebbles, making it impossible to not have pebbles in the disc in the first place. This is self-consistently implemented in \textsc{chemcomp} following \cite{Lenz2019}. We also consider a scenario in which we form planetesimals but do not accrete them to underline how planetesimal formation process reduces the pebble flux and how this reduction influences the composition of the disc and the planet.

\section{Model}
\label{sect:model}
The theoretical assumptions of this model are presented in more detail in \cite{Bitsch2021_1}. 
We use the classic viscous evolution disc model \citep{Lynden-Belle-Pringle1974}, where we solve the viscous evolution equation for each chemical species separately.
We follow the two population approach for dust growth from \cite{Birnstiel2012}, where the full power-law distribution of grain sizes is divided into two bins: small grains that are tightly coupled to the gas and thus are not influenced by drift velocities and large grains that are significantly drifting inwards. The so described dust is then evolved by means of a single advection-diffusion equation using a mass weighted velocity. The planet for computational simplicity accretes the large grain population only. The prescription for pebble accretion originates from \cite{Johansen2017}. The planets grow to pebble isolation mass using the recipe of \citet{Bitsch_2018} while drifting through the disc in type I migration following \cite{Paardekooper_2011}, and \citet{Masset_2017} for the heating torque expression. After reaching pebble isolation mass, the planets have opened a deep enough gap to migrate in type II migration regime. For both the gap opening process and type II migration we follow the recipes in \citet{Ndugu_2021}.
The \textsc{chemcomp} code also includes a routine that allows inward drifting pebbles to evaporate at their corresponding evaporation fronts, resulting in an enrichment of the disc with vapour \citep[e.g.][]{Bitsch2021_1}). We also assume that the original chemical composition does not change due to chemical reactions on the dust grains during the simulation, because the pebble drift time scales are shorter than the chemical reaction time scales \citep[e.g.][]{Booth_2019, Eistrup_2022}.

Following the approach in \citet{Bitsch2021_1}, the planet initially grows by accreting pebbles until it reaches the pebble isolation mass \citep[e.g.][]{Lambrechts_2014,Ataiee_2018,Bitsch_2018}, after which it switches to gas accretion \citep{Ndugu_2021}. During the pebble accretion phase, $90\%$ of the material is attributed to the core, while $10\%$ of the pebbles are attributed to a primordial atmosphere, following other models \citep[e.g.][]{Bitsch2021_1}. We discuss the effect of varying this ratio in Sect. \ref{sect:discussion}. We include a recipe for planetesimal formation from the pebble flux \citep{Lenz2019} based on the idea that planetesimals form in `pebble traps' due to a locally enhanced dust-to-gas ratio (Appendix \ref{app:pla_form}) and consequently planetesimal accretion onto the planets following \citet{Johansen2019}, with an improved capture radius model \citep{Valletta2021}, as  explained in Appendix \ref{app:radius}. In this case, planetesimal accretion can happen at all stages of planet evolution: during core accretion (until pebble isolation mass), the planetesimals are added to core, while during the gas accretion phase accreted planetesimals can pollute the envelope of the planet. The parameters within our models can be found in Appendix~\ref{app:parameters}, while the new implementations for planetesimals (formation and accretion) are described in Appendix \ref{app:pla_form_acc}. Appendix~\ref{app:density} shows the gas, pebble and planetesimal surface density evolution for the scenarios with and without planetesimal formation. Appendix \ref{app:atmo_comp} shows the chemical compositions of the atmospheres (same as Fig. \ref{fig:atm_comp}) for the $10$ and $30$ AU planets, while Appendix \ref{app:growth_tracks} shows the planets' growth tracks and the evolution of their atmospheric C/O ratio. Finally, Appendix \ref{app:heavy} is devoted to describing the origin of the heavy element contents of the planets, originating from pebbles, planetesimals and vapour enriched gas.

\section{Results}
\label{sect:results}
\subsection{Water content of the disc and mass of planetesimals and pebbles}

   \begin{figure*}
   \centering
   \includegraphics[width = \textwidth, keepaspectratio]{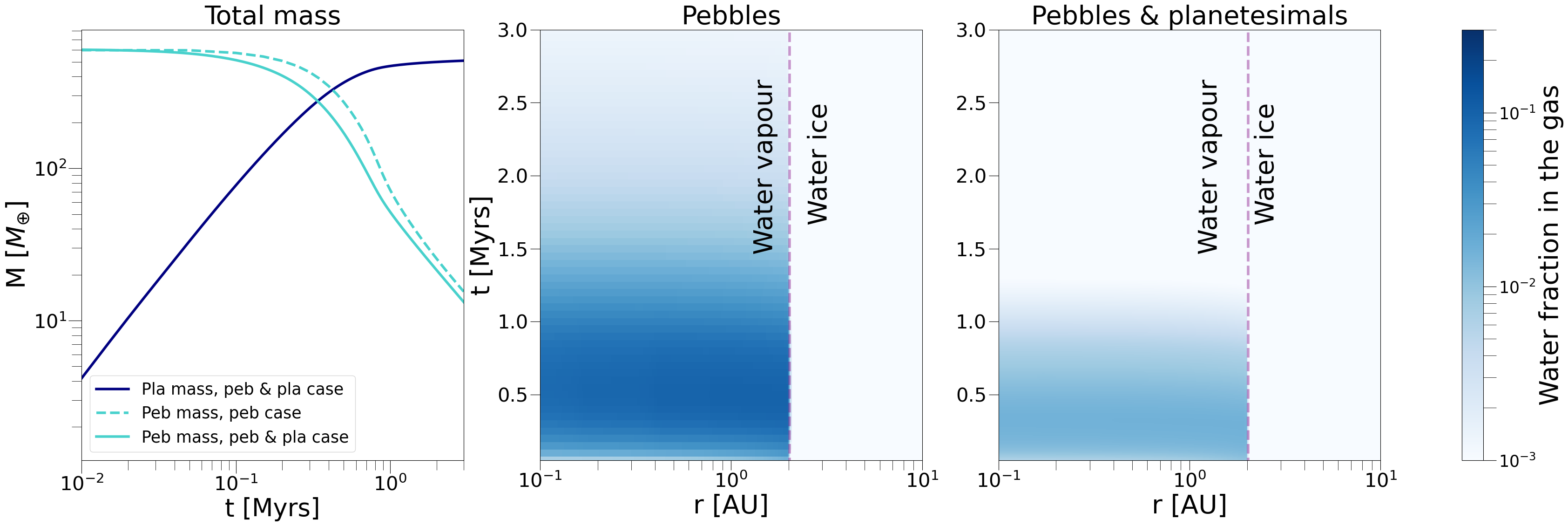}
   \caption{Total mass of pebbles and planetesimal and water fraction in the gas in the disc. Left panel: Total mass of pebbles (light blue lines) and planetesimals (dark blue line) in the two scenarios: pebble accretion only (dotted line) and planetesimal formation (solid lines). Central and right panel: Water content in the gaseous phase of the disc with viscosity $\alpha=10^{-3}$ as a function of radius and time in the case of no planetesimal formation (central panel) and in presence of planetesimal formation (right panel). The vertical violet line marks the water evaporation front in the disc.}
    \label{fig:water_disk}%
    \end{figure*}
The left panel of Fig. \ref{fig:water_disk} shows the total mass of pebbles (light blue lines) and planetesimals (dark blue line) as a function of time for the pebble accretion scenario (dotted line) and the planetesimal formation scenario (solid lines).
The total pebble mass decreases with time in both cases, but reduces faster in the case of planetesimal formation due to fact that the formed planetesimals lock away pebbles (cfr. light blue lines in Fig. \ref{fig:water_disk}).
The planetesimal total mass (solid blue line), instead, increases with time.

The middle and right panels of Fig. \ref{fig:water_disk} show the evolution of the water content of the gas in the disc in time with and without planetesimal formation.
We notice that in both cases at the early stages ($<$ 200 kyr) the water fraction in the gas is low because pebbles still did not have the time to drift inwards and enrich the inner part of the disc with water vapour. As the disc evolves and the pebbles drift, the water content increases. 
In the planetesimal formation case, the water enrichment is clearly limited by the fact that a large number of pebbles are locked into planetesimals, and thus cannot drift inwards, evaporate, and enrich the gas in water vapour. This is showcased for water vapour, but the same reasoning applies to every chemical species that we consider in the simulations (see appendix~\ref{app:parameters}).

\subsection{Atmospheric composition of the planet}
   \begin{figure*}
   \centering
   \includegraphics[width = \textwidth, keepaspectratio]{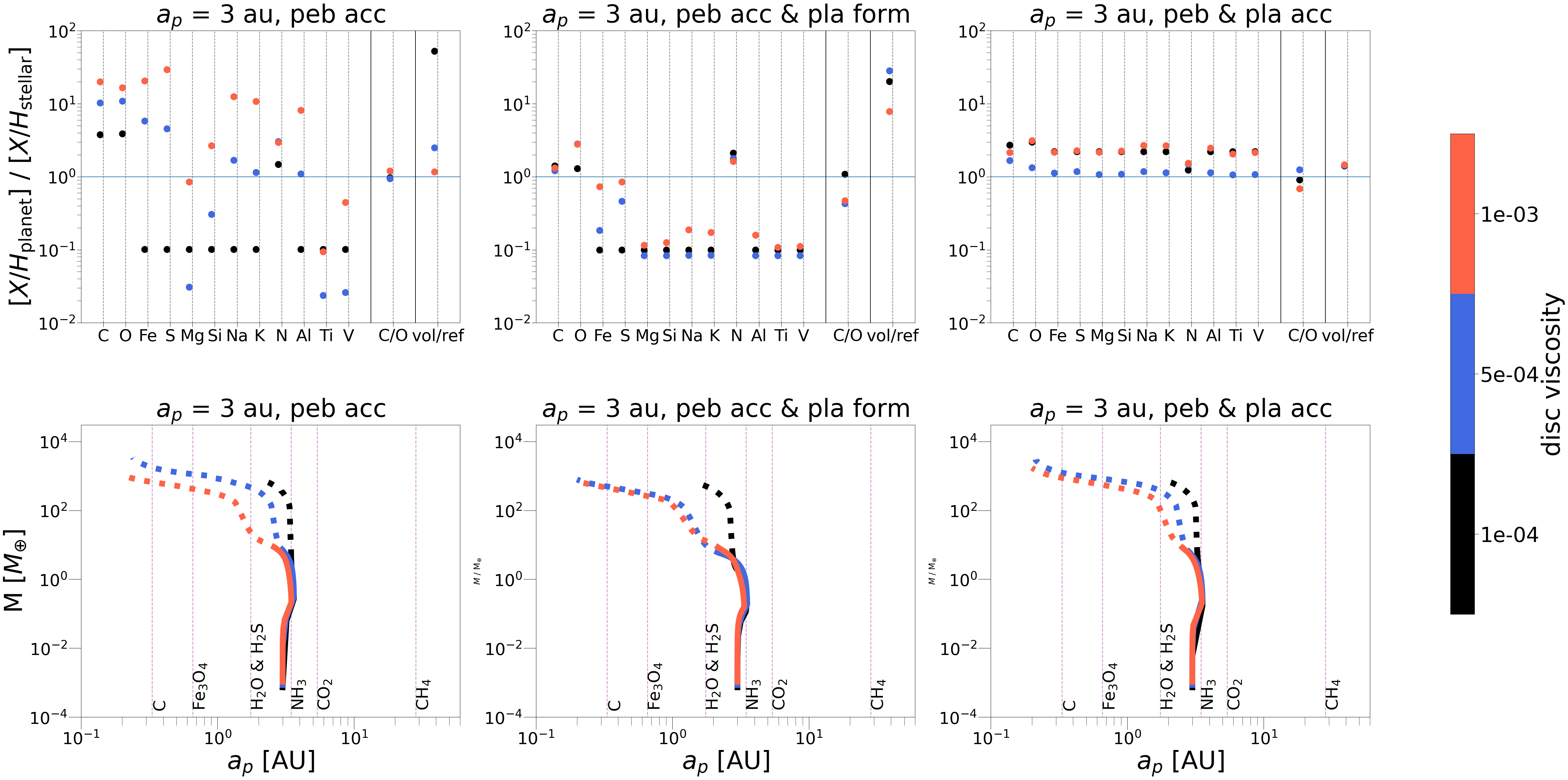}
   \caption{Final elemental abundances of the planetary atmospheres (top) and their corresponding growth tracks (bottom) for the three different scenarios of only pebble accretion (left), planetesimal formation (middle) and pebble and planetesimal accretion (right). The horizontal blue line in the first row marks the solar abundance, while the vertical violet lines in the second row show the evaporation fronts of the chemical species included in our model for a disc viscosity of $\alpha = 5 \cdot 10^{-4}$.The solid lines of the growth tracks correspond to core formation, while the dotted lines correspond to the gas accretion phase. The different colour codings represent different disc viscosities.}
    \label{fig:atm_comp}
    \end{figure*}

We show in Fig.~\ref{fig:atm_comp} the atmospheric composition (top) and the growth tracks (bottom) for planets starting at 3 AU in our three different scenarios (left to right). In particular, we show the normalized abundances as well as the C/O ratio and the volatile to refractory ratio, where species with $T_{\mathrm{cond}} \leq 150$ K are considered volatiles and species with $T_{\mathrm{cond}} > 150$ K are considered refractories \citep{Bitsch2021_1,Bitsch2021_2}. The different colours refer to different disc viscosities. The bottom row shows the corresponding growth tracks. We show results for planets starting at 10 and 30 AU in Appendix \ref{app:atmo_comp}.

In the pebble accretion only scenario (left column), the planets have clearly super-solar C/H and O/H ratios, because the drifting pebbles efficiently enrich the gas in volatile content that is subsequently accreted onto the planet. The different viscosities act on the composition of the planets in two different ways: larger viscosities result in a faster migration of the planet that, therefore, crosses more evaporation fronts and is able to accrete enriched gas of species that are not available in gaseous form for the slower migrating planet at low viscosity. 
The total enrichment of the atmosphere then crucially depends on which evaporation fronts are crossed by the growing planet.
However, at higher viscosities the gas is less enriched in volatiles because the gas transport is faster \citep[see][]{Bitsch2021_1, Mah_2023}.

If we introduce planetesimal formation in the disc (middle column), the planets grow slightly less massive due less solids available to grow their cores. A general depletion of the elemental abundances with respect to the pure pebble accretion case is observed, because of the locking of pebbles into planetesimals. This depletion is more significant for higher viscosities (red dots) because they are overall the most enriched planets in the pebble accretion case, resulting therefore in a bigger depletion when the disc is less enriched.
We observe an increase in the volatile to refractory ratio in the case of planetesimal formation because of the depletion in the refractories locked into planetesimal\footnote{This is caused by the fact that the planetesimal surface density is very steep function with radius, proportionally locking more pebbles into planetesimals interior of the water ice line compared to exterior to the ice line (Fig. \ref{fig:surface_density}).} that in this scenario are not accreted onto the planet.

   \begin{figure*}
   \centering
   \includegraphics[width = \textwidth, keepaspectratio]{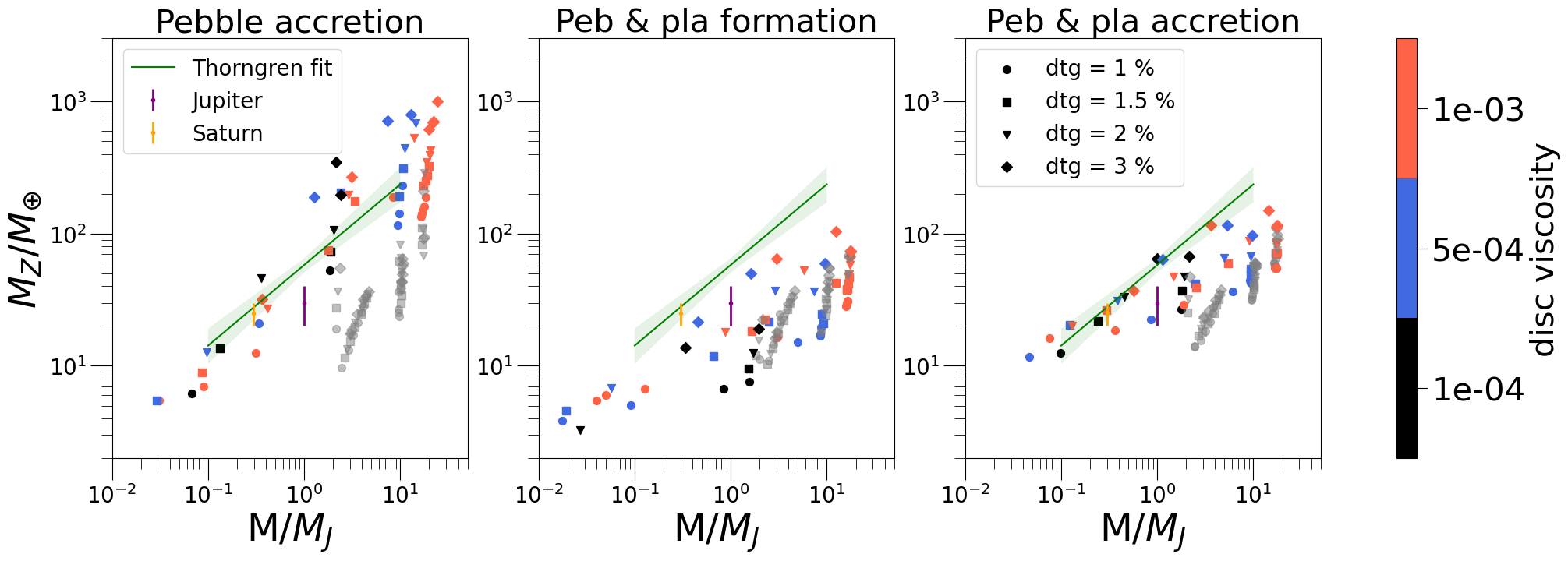}
   \caption{Total heavy element content of the planets with final mass $M>5M_{\oplus}$ and position $a_{\mathrm{p}} < 1$ AU as a function of the total mass for the three formation scenarios. The colour coding represents the different viscosities, while the different markers showcase different initial dust-to-gas ratios of the disc. The green line is the fit from \citet{Thorngren2016}, while Jupiter and Saturn are marked in purple and orange respectively. The grey points represent planets that end up with $a_{\mathrm{p}} > 1$ AU from the central star.}
              \label{fig:heavy_elem}%
    \end{figure*}

The last scenario (right column) shows planets formed through pebble and planetesimal accretion. In this case we observe a significant increase of refractories and volatiles due to the accretion of planetesimals compared to the scenario of only planetesimal formation.
Interestingly, independently of the formation scenario, the final atmospheric C/O ratio is largely unaffected by the formation method, even though the evolution of the atmospheric C/O ratio changes within the different formation methods, see Appendix \ref{app:growth_tracks}.

\citet{Bitsch2021_1} suggested that the volatile to refractory ratio of atmospheres could be used to distinguish between the different accretion scenarios \citep[see also][]{Chachan_2023, Knierim_2022}. Generally, the C/H and O/H ratio of planets formed in the pure pebble scenario are larger compared to the scenario with planetesimal formation and accretion. However, the accretion of refractory rich planetesimals leads to a low volatile to refractory ratio. However, also the pure pebble scenario can produce planets with low volatile to refractory ratio, if they migrate all the way to the inner disc, where also refractories evaporate. Distinguishing the different scenarios now requires additionally a detailed measurement of C/H or O/H and not only the volatile to refractory ratio, because C/H and O/H are much larger in the pure pebble scenario compared to the planetesimal scenario (e.g. compare the planets marked in red). 
This could therefore be a tracer for the formation pathway of a planet.

\subsection{Planet's heavy element content and atmospheric metallicity}
 Figure \ref{fig:heavy_elem} shows the total heavy element content of the planets formed in our different sets of simulations.
 The green line shows the fit from \citet{Thorngren2016}, although a more recent analysis \citep{Bloot_2023} seems to highlight a lower heavy element content for planets below $2 M_{J}$ masses by taking also constraints from atmospheric measurements into account.
It can be clearly noticed that there is a significant difference in the heavy element content of the planets created with the pebble accretion model with respect to the other two scenarios.
As explained above, if planetesimals form in the disc and are not accreted by the planet, the total heavy element content of the planet drops because material is locked into them. 
Even when planetesimal accretion is allowed, the heavy element content of the planets stays much lower compared to the pure pebble scenario, in line with \citet{Venturini_2020}.
This indicates that planets with large heavy element content are most likely born in discs where planetesimal formation is inefficient and should consequently harbour larger C/H and O/H values, testable via observations.

\begin{figure}
   \centering
   \includegraphics[width = 0.5
\textwidth, keepaspectratio]{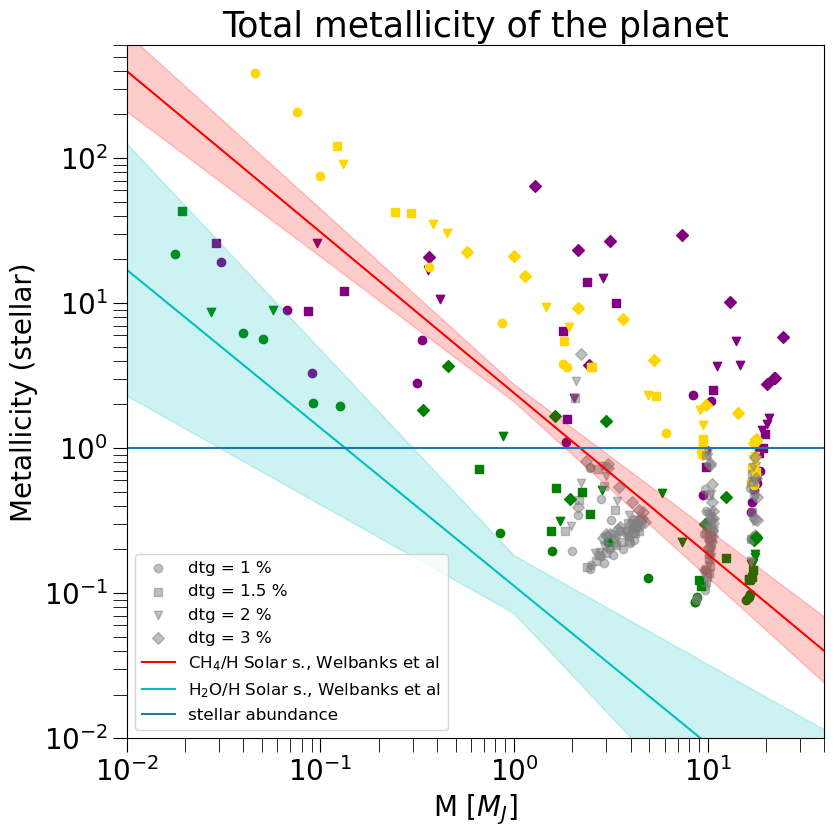}
   \caption{Atmospheric metallicity as function of planetary mass for the different formation scenarios (purple = pebble accretion, green = planetesimal formation, gold = pebble and planetesimal accretion). The different markers represent different dust-to-gas ratios and the grey symbols mark planets with $a_{\mathrm{p}} > 1$.}
              \label{fig:metallicity_planet}%
    \end{figure}

Figure \ref{fig:metallicity_planet} shows the total metallicity of the simulated planets compared to the stellar metallicity as a function of planetary mass. The pebble accretion scenario (purple dots) generates planets with the highest metallicity for final masses above 1 Jupiter mass, while for planets with $M < 1 M_{\mathrm{J}}$ the highest metallicity is found in the combined pebble and planetesimal accretion scenario (gold dots). Even though these planets have the highest atmospheric metallicity, their total heavy element content is similar to the planets formed in the pure pebble scenario (Fig. \ref{fig:heavy_elem}). The difference arises from the fact that the slower pebble accretion rate in the planetesimal scenario allows the planets to migrate inwards further compared to the pure pebble scenario, however, in the inner disc the pebble isolation mass is smaller due to the lower aspect ratio \citep{Bitsch_2015a}, resulting in lower core masses of these planets. Consequently, these planets have a larger atmospheric metallicity if they have the same heavy element mass as the planets formed in the pure pebble scenario.
The planets that form in the planetesimal formation scenario (green dots) have the lowest metallicity as expected.

It is striking to observe that nearly all the planets whose final location is beyond 1 AU (grey dots) have sub-stellar atmospheric metallicity, while the inner ones are mainly super-stellar. This implies that planets with sub-stellar atmospheric metallicity form in the outer disc, exterior to the main evaporation fronts (see discussion in \citealt{Bitsch2021_1,Bitsch_2022}). Thus, if we observe for example hot-Jupiters with sub-stellar atmospheric metallicity, it means that they probably formed in the outer disc and underwent a scattering event that brought them to closer orbits to the central star. Hot-Jupiters with super-solar metallicity, instead, are mostly migration driven. In addition, planets formed in discs with larger metallicities are more enriched in heavy elements, as expected \citep{Bitsch2021_1}.

\section{Model limitations}
\label{sect:discussion}
The pebble evolution and accretion has been simulated using a constant fragmentation velocity of $5$ m/s, following laboratory experiments that did not find differences in the fragmentation velocity between silicates and water ice \citep{Musiolik_2019}.
Higher fragmentation velocities would lead to bigger sized pebbles that in turn would migrate faster inwards, making pebble accretion more efficient, while lower velocities will result in smaller pebbles that would drift inwards on longer time-scales eventually prolonging the planet formation process and thus still allowing the formation of giant planets \citep[][]{Savvidou_2023}. The heavy element content can be expected to be initially larger for higher fragmentation velocities because of the faster pollution of the gas phase due to faster inward drifting of pebbles, however it also declines faster for large disc viscosities.

The planet's envelope opacity is a key parameter for the contraction and gas accretion phase, as well as for the planetesimal accretion radius. 
A low opacity results in a fast gas accretion and therefore an earlier transition to type II migration regime.
In this work we used a fixed value for the opacity consistent with \citet{Movshoivitz_2008}, but we also analysed the effects that changing envelope opacity has on the planetesimal accretion radius (see Appendix \ref{app:radius}).
A higher envelope opacity allows the planet to stay for
a longer time in the attached phase, reaching the constant planetesimal capture radius
at a later time. Consequently, the planets could be enriched with more planetesimals, because the planets feature a larger capture radius for a longer time.

The planetesimal formation model used in this work follows \citet{Lenz2019} and is based on the idea that planetesimals can form at any location, as long as pebbles are available. We used this model because we wanted to analyse the limiting case in which we have a large  planetesimal population. 
We used a fixed planetesimal formation efficiency parameter according to \citet{Lenz2019}, but we also tested for different efficiencies. Larger formation efficiencies lead to stronger depletions in the pebble surface density, resulting in a less efficient pebble accretion and even potentially hindering it \citep[see also][]{Kessler_2023}.
Too low planetesimal formation efficiencies would, instead, lead back to the pebble accretion only scenario.
We chose, therefore, a value for the efficiency that was sufficiently large to easily form planetesimals but not too large to prevent pebble accretion.
The planetesimal formation model of \citet{Drazkowska2017} predicts planetesimal formation only around the water ice line. Consequently, planets forming completely exterior to the water ice line would not be affected by planetesimal formation, while planets forming interior to the water ice line would harbour reduced metallicities compared to the pure pebble scenario. In addition, this planetesimal formation scenario would open the questions how giant planets could accrete refractory materials without migrating into the very inner disc.

The dust is evolved using the two population approach from \citet{Birnstiel2012}, that divides the full power-law distribution of dust grains into two size bins: the small population, which is the part of the size distribution which is not influenced by drift velocities because the particles are small enough to be tightly coupled to the gas and the large population, which are the grains that are significantly drifting inwards. This approach is clearly a simplified treatment of the dust size distribution that we can observe in protoplanetary discs, but has the advantage of being computationally fast, making it feasible to perform many simulations while still giving rather accurate results \citep[e.g.][]{Andama_2022, Stammler_2023}.

The planetesimal accretion scenario considers just one size of planetesimals, in agreement with other works of planet formation via planetesimal accretion \citep[e.g.][]{Emsenhuber_2022}. Furthermore, as shown in Fig. \ref{fig:capture_radius}, the actual size of accreted planetesimal produces in our case a small difference, meaning that considering a full size distribution of planetesimal would not change our results significantly but would increase the computational complexity of the model.

An important assumption of this model is that during the initial phase of pebble accretion $10\%$ of the accreted material builds up a primordial atmosphere\footnote{This value originates from the fact that the envelope of Uranus and Neptune contributes roughly to 10\% of their total mass.} \citep[e.g.][]{Bitsch2021_1}. This is a simplified way of treating the problem that accreted particles sublimate during the core build-up phase. More sophisticated models that take into account the structure of the envelope show that up to $50\%$ of the initially accreted pebbles could form a primordial atmosphere \citep{Brouwers_2021}.
While clearly more sophisticated approaches are needed to understand the accretion of heavy elements onto growing giant planets during the core growth phase, our general trends would not be affected by this. The reason is that the cores in all our scenarios are mainly formed through pebble accretion (due to the inefficiency of planetesimal accretion at large distances, e.g. \citet[][]{Johansen2019}), implying that they should have the same heavy element content due to evaporated pebbles. While it is clear that the absolute value of enrichment might change, the general trend that the pure pebble accretion scenario allows larger total heavy element contents will not change.
On the other hand, a larger primordial heavy element envelope that is then mixed with the atmosphere of the planet might influence the atmospheric C/O ratio. Nevertheless, the overall trend that planets forming further away from the star harbour larger C/O ratios will remain intact, because their heavy element mass originates mainly from gas and planetesimals rather than from pebble accretion, which happens only during the core formation stage, as we show in Appendix~\ref{app:heavy}.

We make the assumption that the atmospheres are evenly mixed, as for hot Jupiters \citep[e.g.][]{Guillot_2022}. However, this is not true for Jupiter in our own solar system, where compositional gradients exist \citep[e.g.][]{Wahl_2017,Vazan_2018}.

\section{Summary and conclusions}
\label{sect:conclusion}
We performed 1D semi-analytical simulations of growing planets in a protoplanetary disc, tracing their chemical composition, using the \textsc{chemcomp} code \citep{Bitsch2021_1}.
We considered three different formation scenarios: planetary growth through pure pebble accretion, growth through pebble accretion with the possibility of forming planetesimals in the disc but not accreting them on the planet, and combined growth by pebble and planetesimal accretion.
In all scenarios the starting embryo accretes pebbles until it reaches the pebble isolation mass, then switches to gas accretion. In the combined growth scenario the embryo can also accrete planetesimals throughout its entire life, allowing extra solids to be accreted and added to the atmosphere.

Our simulations show that planetesimal formation strongly reduces the volatile enhancement in the disc that is caused by pebble drift and evaporation, (see Fig. \ref{fig:water_disk}).
Consequently, the heavy element content of the grown giant planets is largest in the pure pebble scenario, while it drops if planetesimal formation becomes efficient. Even the additional accretion of planetesimals does not allow to form planets largely enriched in heavy elements in our scenario.
This indicates that planets with high heavy element content are predominantly formed in discs where planetesimal formation is inefficient.

The final atmospheric C/O ratio of the planets depends on the final mass of the planet and how and when it migrates through the disc and the corresponding evaporation fronts and is different for the three scenarios. 
Generally we do not find a pattern in the C/O ratio that allows us to distinguish the different formation scenarios. Thus, we conclude that the C/O ratio alone is not a good tracer to distinguish the different formation scenarios \citep[see also][]{Bitsch_2022, Molliere_2022}.

Our simulations show that planetesimal formation might hinder the enrichment of planetary atmospheres compared to the pebble accretion scenario, but can provide a low volatile-to-refractory ratio in contrast to the pure pebble scenario, unless, the planet migrates into the inner region of the disc, where also refractories evaporated and could be accreted with the gas. The differences in planetary compositions are large enough that future observations could distinguish between the different formation channels, allowing further constraints to planet formation models.

\begin{acknowledgements}
      C.D. thanks the European Research Council (ERC Starting Grant 101041466-EXODOSS) for their financial support and Marten B. Scheuck and Molly R. A. Wells for helpful discussion, B.B. thanks the European Research Council (ERC Starting Grant 757448-PAMDORA) for their financial support. J.M. and B.B. acknowledge the support of the DFG priority program SPP 1992 ``Exploring the Diversity of Extrasolar Planets'' (BI 1880/3-1).
      We thank an anonymous referee for her/his report that helped to improve the quality of this manuscript.
\end{acknowledgements}

\bibliographystyle{aa} 

\begin{appendix}
\section{Parameters used for simulations}
\label{app:parameters}

Table \ref{tab:param_disc} shows the set of disc parameters used in the simulations, while Table \ref{tab:param_planet} shows all the possible planet parameters used in the simulations. In Table~\ref{tab:chemical_species} we show our chemical partitioning model, where we put 60\% of the carbon into refractory carbon grains. The detailed chemical partitioning influences the detailed composition of the formed planets \citep{Bitsch2021_1}, but has only a tiny influence on the heavy element content of forming giant planets.

\begin{table}[h]
\centering
    \caption{Disc parameters}
    \begin{tabular}{ ccc }
        \hline
        \hline
        Quantity & Value & Meaning\\
        \hline
        $\alpha$  & $[1, 5, 10] \cdot 10^{-4}$  &  alpha viscous parameter \\
        $\alpha_z$ &  $1 \cdot 10^{-4}$   &  vertical mixing \\
        $M_0$ & $0.128 M_{\odot}$ & initial disc mass\\
        $R_0$ & $137$ AU & initial disc radius \\
        $t_{\mathrm{evap}}$ & 3 Myr & disc lifetime \\
        $\mathrm{dtg}$ & $[1 \%, 1.5 \%, 2 \%, 3\% ]$ & dust-to-gas ratio \\
        $u_{\mathrm{frag}}$ & $5$ m/s & fragmentation velocity \\
        $R_{\mathrm{pla}}$ & $1$ km & radius of planetesimals \\
        \hline
    \label{tab:param_disc}

    \end{tabular}
\end{table}

\begin{table}[h]
\centering
    \caption{Planet parameters}
    \begin{tabular}{ ccc }
        \hline
        \hline
        Quantity & Value & Meaning\\
        \hline
        $a_p$  & [1, 2, 3, 5, 10 & initial position of embryo\\
         & 15, 20, 25, 30] AU  &   \\
        $t_0$ &  0.05 Myr   &  implantation time of embryo \\
        $\kappa_{\mathrm{env}}$ & 0.05 $\mathrm{cm^2 g^{-1}}$ & envelope opacity\\
        \hline
        \label{tab:param_planet}

    \end{tabular}
\end{table}

\begin{table*}
\centering
    \caption{Condensation temperatures and volume mixing ratios of chemical species treated in the code.}
    \begin{tabular}{ccc}
        \hline\hline
        Species (Y)  & $T_{\mathrm{cond}} [K]$ & $v_{Y}$\\
        \hline
        $\mathrm{CO}$  & 20  & $0.2 \cdot \mathrm{C/H}$\\
        $\mathrm{N_2}$ &  20   &  $0.45 \cdot \mathrm{N/H}$ \\
        $\mathrm{CH_4}$ & 30 & $0.1 \cdot \mathrm{C/H}$ \\
        $\mathrm{CO_2}$ & 70 &   $0.1 \cdot \mathrm{C/H}$\\
        $\mathrm{NH_3}$ & 90 & $0.1 \cdot \mathrm{N/H}$ \\
        $\mathrm{H_2S}$ & 150 & $0.1 \cdot \mathrm{S/H}$ \\
        $\mathrm{H_2O}$ & 150 & $\mathrm{O/H} - (3 \cdot \mathrm{MgSiO_3/H} + 4 \cdot \mathrm{Mg_2SiO_4/H} + \mathrm{CO/H}$ \\
         & & + $2 \cdot \mathrm{CO_2/H} + 3 \cdot \mathrm{Fe_2O_3/H} + 4 \cdot \mathrm{Fe_3O_4/H} + \mathrm{VO/H} $ \\
          & & + $\mathrm{TiO/H} + 3 \cdot \mathrm{Al_2O_3/H} + 8 \cdot \mathrm{Na Al Si_3 O_8/H} + 8 \cdot \mathrm{KAlSi_3O_8/H})$ \\

        $\mathrm{Fe_3O_4}$ & 371 & $(1/6) \cdot (\mathrm{Fe/H - 0.9 \cdot \mathrm{S/H}})$\\
        $\mathrm{C}$ (carbon grains) & 631 & $0.6 \cdot \mathrm{C/H}$  \\
        $\mathrm{FeS}$ & 704 & $0.9 \cdot \mathrm{S/H}$  \\
        $\mathrm{NaAlSi_3O_8}$ & 958& $ \mathrm{Na/H}$\\
        $\mathrm{KAlSi_3O_8}$ & 1006 & $ \mathrm{K/H}$ \\
        $\mathrm{Mg_2SiO_4}$ & 1354 &$\mathrm{Mg/H} - (\mathrm{Si/H} - 3 \cdot \mathrm{K/H} - 3 \cdot \mathrm{Na/H})$\\
        $\mathrm{Fe_2O_3}$ & 1357 & $0.25 \cdot (\mathrm{Fe/H - 0.9 \cdot \mathrm{S/H}})$\\
        $\mathrm{VO}$ & 1423 & $\mathrm{V/H}$\\
        $\mathrm{MgSiO_3}$ & 1500 & $\mathrm{Mg/H} - 2 \cdot (\mathrm{Mg/H} - (\mathrm{Si/H} - 3 \cdot \mathrm{K/H} - 3 \cdot \mathrm{Na/H}))$ \\
        $\mathrm{Al_2O_3}$ & 1653 & $0.5 \cdot (\mathrm{Al/H} - (\mathrm{K/H} + \mathrm{Na/H}) )$ \\
        $\mathrm{TiO}$ & 2000 & $ \mathrm{Ti/H}$ \\
        \hline
        \label{tab:chemical_species}
    \end{tabular}
\end{table*}

\section{Planetesimal formation and accretion}
\label{app:pla_form_acc}
\subsection{Planetesimal formation}
\label{app:pla_form}
In this work we use the planetesimal formation model presented in \cite{Lenz2019}, based on the idea that planetesimals form in `particle traps' that can emerge everywhere in the disc resulting in a local enhancement of the dust-to-gas ratio, allowing the formation of planetesimals.
The column density of drifting pebbles can be converted into column density of planetesimals by means of:
\begin{equation}
    \Dot{\Sigma}_{\mathrm{pla}} (r) = \frac{\varepsilon}{d(r)} \frac{\Dot{M}_{\mathrm{peb}}}{2 \pi r},
\end{equation}
where $\varepsilon$ is the efficiency parameter, $r$ the distance from the central star, $d(r)$ the radial separation of pebble traps and $\Dot{M}_{\mathrm{peb}}$ the pebble flux given by 
\begin{equation}
    \Dot{M}_{\mathrm{peb}} = 2 \pi r \sum_{\mathrm{St_{min} \leq St \leq St_{\mathrm{max}}}} |v_{\mathrm{drift}}(r, \mathrm{St})|\Sigma_{\mathrm{d}}(r,\mathrm{St}),
\end{equation}
where $v_{\mathrm{drift}}$ is the radial drifting velocity of the particles, $\Sigma_{\mathrm{d}}(r,\mathrm{St})$ the column density (in particles) that have the required Stokes number and $\mathrm{St_{min}}$ and $\mathrm{St_{max}}$ are the minimum and maximum Stokes number of particles that are able to participate in the streaming instability to facilitate gravitational collapse and planetesimal formation.
We do not use a limiting Stokes number in agreement with \citet{Lenz2019}. We want to point out that our Stokes numbers in the outer disc are always larger than $10^{-3}$, due to the large fragmentation velocity threshold. Particles with Stokes numbers of $10^{-3}$ can already contribute to the streaming instability \citep{Line_2021}, in agreement with our Stokes numbers.

\subsection{Planetesimal accretion}
To model the planetesimal accretion rate we follow \citet{Johansen2019}, based on the model of \citet{Tanaka_1999}.
The planetesimal accretion rate, in the case of a single migrating protoplanet, is given by:
\begin{equation}
\label{eqz:acc_rate}
    \Dot{M} = \epsilon_{\mathrm{pla}} \Dot{M}_{\mathrm{pla}} = \epsilon 2 \pi r \Dot{r} \Sigma_{\mathrm{pla}},
\end{equation}
where $\epsilon_{\mathrm{pla}}$ is the accretion efficiency and $ \Dot{M}_{\mathrm{pla}}$ is the flux of planetesimals that cross the orbit of the migrating protoplanet. The migration speed is computed using the inverse of the normalized migration timescale:
\begin{equation}
    \Dot{\Tilde{b}}_{\mathrm{p}} = \Tilde{\tau}_{\mathrm{mig}}^{-1},
\end{equation}
while the accretion efficiency is given by:
\begin{equation}
    \epsilon_{\mathrm{pla}} = \alpha_{\mathrm{pla}}\Dot{\Tilde{b}}_{\mathrm{p}}^{\beta_{\mathrm{pla}-1}},
\end{equation}
with $\alpha_{\mathrm{pla}}$ and $\beta_{\mathrm{pla}}$ being fits to numerical simulations given by \citep{Tanaka_1999}:
\begin{align}
\label{eqz:parameters}
\alpha_{\mathrm{pla}} &= 2.5\sqrt{\frac{\Tilde{R}_{\mathrm{p}}}{1+0.37\Tilde{i}_0^2/\Tilde{R}}_{\mathrm{p}}},\\
\beta_{\mathrm{pla}} &=0.79(1+10 \Tilde{i}^2_0)^{-0.17},
\end{align}
where $\Tilde{i}_0$ is the inclination of the planetesimal population, $\Tilde{R}_p$ the planetesimal accretion radius normalized to the Hill radius.
We use here a vertical stirring of the planetesimals of $\delta_{\mathrm{stir}}=10^{-4}$ to calculate the inclination of the planetesimal distribution following \citet{Ida_2008}.

\subsection{Capture radius}
\label{app:radius}
The capture radius for planetesimal accretion is modelled according to \citet{Valletta2021}. In principle, in order to correctly predict the capture radius for planetesimal accretion, one needs to compute the trajectories of the planetesimals in the disc, solve the stellar structure equation within the protoplanet's envelope and take into account the drag force that the atmosphere exerts on the planetesimals. 
\citet{Inaba2003} have shown that the capture radius is significantly larger than the core's radius and that it depends mostly on the planetesimal size: the smaller the planetesimal, the bigger the capture radius.
\citet{Valletta2021} propose two different prescription for the capture radius, depending on whether the planet is still embedded in the disc (attached phase) or not (detached phase).
During the attached phase, the planet's temperature and density are determined by its orbital location. The outer radius of the planet is defined as:
\begin{equation}
    R_{\mathrm{p, out}} = \frac{GM_{\mathrm{p}}}{c_{\mathrm{s}}^2 + \frac{GM_{\mathrm{p}}}{0.25 R_{\mathrm{H}}}},
\end{equation}
with $G$ being the gravitational constant, $M_{\mathrm{p}}$ the planet mass, $R_{\mathrm{H}}$ the Hill radius, and $c_{\mathrm{s}}$ the sound speed in the disc.
The capture radius in this phase depends on the gas drag on the planetesimals, which is itself affected by the density profile of the planet's envelope, meaning that, to determine the capture radius, $R_{\mathrm{capt}}$ an estimate of the planet atmospheric profile is needed. This is achieved using the mass conservation, hydrostatic balance, thermal gradients, and energy conservation equations that regulate the envelope’s structure:
\begin{eqnarray}
    \frac{\mathrm{d}m}{\mathrm{d}r} &=& 4 \pi r^2 \rho, \\
    \label{eqz:pressure}
    \frac{\mathrm{d}P}{\mathrm{d}r} &=& - \frac{Gm}{r^2} \rho, \\
    \frac{\mathrm{d}T}{\mathrm{d}r} &=& \nabla \frac{T}{P} \frac{\mathrm{d}P}{\mathrm{d}r}, \\
    \frac{\mathrm{d}L}{\mathrm{d}r} &=& 4 \pi r^2 \rho \left(\varepsilon -T \frac{\partial S}{\partial t}\right),
\end{eqnarray}
where $m, r$ are the mass and radius coordinate, $\rho, P, T$ are the density, pressure and temperature in the envelope, $L, S$ the luminosity and entropy, and $\nabla = \mathrm{d}\ln T/\mathrm{d}\ln P$ temperature gradient.

In the outer layers of the planet's envelope, radiation transports the heat, resulting in an almost constant temperature profile and an exponentially increasing pressure and density profile towards the centre of the planet:
\begin{eqnarray}
\label{eqz:prof}
    T(r) &=& T_0, \\
    P(r) &=& P_0 \exp{\left[\delta(R_{\mathrm{p, out}}/r-1)\right]}, \\
    \label{eqz:density}
    \rho(r) &=& \rho_0 \exp{\left[\delta(R_{\mathrm{p, out}}/r-1)\right]}.
\end{eqnarray}
By substituting Eq. (\ref{eqz:prof}) into Eq. (\ref{eqz:pressure}), $\delta$ yields:
\begin{equation}
    \delta = \frac{GM\rho_0}{P_0 R_{\mathrm{p, out}}}.
\end{equation}
Now, assuming $m=M$ to be the total mass of the planet, which is a reasonable guess for the outer layers of the planet's atmosphere, we can use Eq. (\ref{eqz:density}) to infer the density profile of the envelope.

At this point, the approximation for the capture radius is obtained by inserting Eq. (\ref{eqz:density}) into Eq. (18)\footnote{\citet{Inaba2003} define $r_{\mathrm{p}}$ as \begin{equation*}
    r_{\mathrm{p}} = \frac{3}{2} \frac{\rho(R_{\mathrm{c}})}{\rho_{\mathrm{p}}} R_{\mathrm{H}},
\end{equation*}
with $R_{\mathrm{H}}$ Hill radius, $R_{\mathrm{c}}$ core radius, $\rho(R_{\mathrm{c}})$ gas density at core radius, $\rho_{\mathrm{p}}$ material density of the planetesimal.
} of \citet{Inaba2003}, obtaining:
\begin{equation}
    \label{eqz:attached}
    R_{\mathrm{capt}} = \frac{R_{\mathrm{p, out}}}{1+\frac{1}{\alpha} \ln{\left(\frac{\rho_*}{\rho_0}\right)}}
\end{equation}
with
\begin{equation}
    \rho_* = \frac{2r_{\mathrm{p}}\rho_{\mathrm{p}}}{3DR_{\mathrm{H}}},
\end{equation}
where $r_{\mathrm{p}}, \rho_{\mathrm{p}}$ are the planetesimal's size and density, and $D$ is the drag coefficient present in Eq. (11) of \citet{Inaba2003}.

Equation (\ref{eqz:attached}) that we derived for the attached phase is no longer valid when the planets run out of the gas supply from the disc and, as a result, detaches from it.
The assumption that we make is that this phase starts when the total mass of helium and hydrogen equals the heavy element mass (this is called the crossover mass), that is a phase in which the planetary radius collapses rapidly and then decreases slowly over time. At the crossover mass, the capture radius can be approximated as a constant, and it depends on the ratio between the heavy element mass and helium-hydrogen mass rather than on the runaway gas accretion rate. 
The planet's capture radius in the detached phase is better represented by the following numerical fit \citep{Valletta2021},
\begin{equation}
    \label{eqz:detached}
    R_{\mathrm{capt}} = \left(\sum_{i = 0}^{4} R_i \frac{M_{\mathrm{Z}}^i}{M^i_{\mathrm{H-He}}}\right) \cdot 10^9 \mathrm{cm},
\end{equation}
where the fit parameters are: $R_0 = 12.80662188 , 9.15426162$, $R_1 = -50.86303789,$ $-6.74548399$, $R_2 = 382.66267044$ , $9.40271959 $, $R_3 = -1388.57741163$ , $0 $ and $R_4 = 1902.60362959$ , $0$. Each of the two values represent the fit after $10^7$ and $10^8$ years of evolution, while the radius between these years can be derived with a logarithmic interpolation between the two values. We use here the value at $10^7$ years.

\begin{figure}
\centering
    \includegraphics[width=\columnwidth]{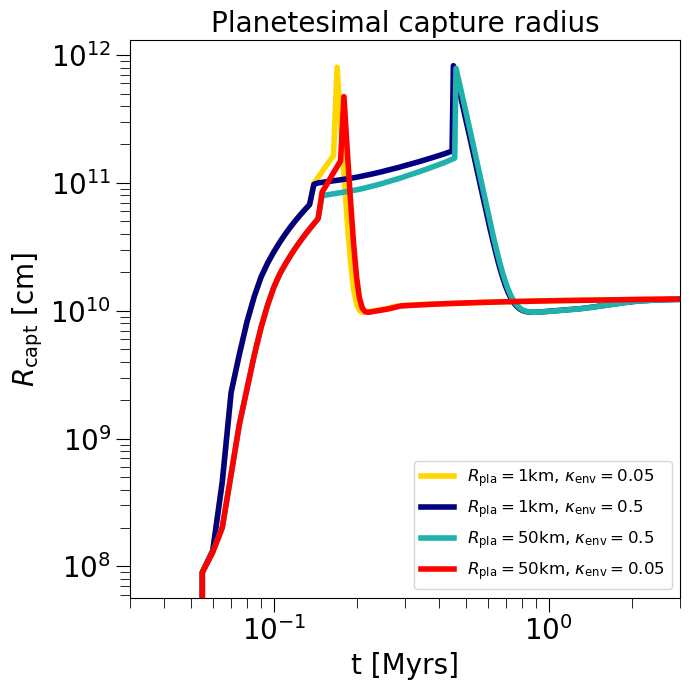}
    \caption{Capture radius as a function of time for $R_{\mathrm{pla}} = 1$ km, $\kappa_{\mathrm{env}} = 0.05$ (gold), $R_{\mathrm{pla}} = 1$ km, $\kappa_{\mathrm{env}} = 0.5$ (navy), $R_{\mathrm{pla}} = 50$ km, $\kappa_{\mathrm{env}} = 0.05$ (red) and $R_{\mathrm{pla}} = 50$ km, $\kappa_{\mathrm{env}} = 0.5$ (light sea green). The simulation shows a non-migrating planet at $3$ AU with a dust-to-gas ratio of $1.5\%$.
    Higher envelope opacities (navy and light sea green lines) lead to a planet that stays longer in the attached phase.}
    \label{fig:capture_radius}
\end{figure}

We ran simulations to study the planetesimal capture radius for different planetesimal sizes and envelope opacity.
Figure \ref{fig:capture_radius} shows the planetesimal capture radius we obtained following Eqs. (\ref{eqz:attached}) and (\ref{eqz:detached}).
We observe a weak dependence of the accretion radius on the planetesimal size in the attached phase and independence (by definition) in the detached phase. All the results are in good accordance with \citet{Valletta2021}.
The size of the accreted planetesimals seems not to have a significant impact on the final heavy element mass of the planets,  because the capture radius is not altered significantly.

\section{Gas and solid surface densities}
\label{app:density}

   \begin{figure*}
   \centering
   \includegraphics[width = \textwidth, keepaspectratio]{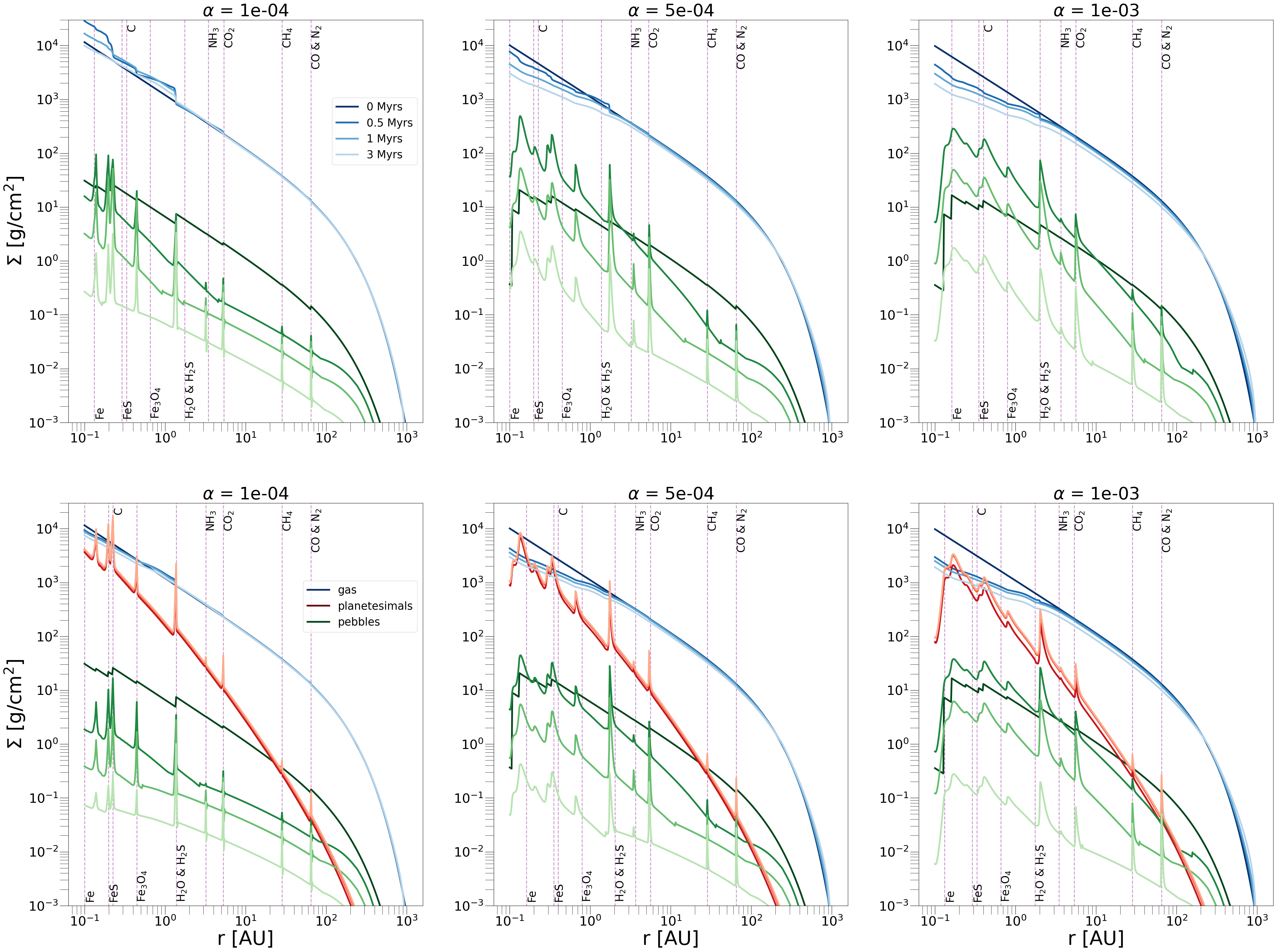}
   \caption{Surface densities of gas, pebbles and planetesimals for the disc described in Table \ref{tab:param_disc} in absence of planets, for different disc viscosities increasing from left to right. The top panel shows the pebble accretion scenario, where planetesimals cannot form, the bottom panel shows what happens, instead, when planetesimal formation is involved.}
            \label{fig:surface_density}%
    \end{figure*}
Figure \ref{fig:surface_density} shows the gas, pebble\footnote{We recall that the pebble surface density is obtained from the dust surface density multiplying it by a factor $f_m$.} and planetesimal surface densities of discs with different viscosities as a function of radius and time, in the pebble accretion only scenario (top panel) and in presence of planetesimal formation (bottom panel). The vertical dotted lines represent the evaporation fronts of some molecules that we consider in our model.

We observe the same trend of time evolution of gas and pebble surface densities. 
In both cases, we observe the gas surface density (blue line) to decrease with time in the inner part of the disc, due to the accretion onto the protostar.

The pebble surface density (green line) in both cases shows spikes at the evaporation lines, due to the fact that immediately exterior to the evaporation line, the gas re-condenses into dust forming new pebbles, thus increasing the local pebble surface density. Furthermore, it first increases with time in the inner disc, then it decreases as pebbles are used either to form planets or drift into the central star.
The pebble surface density shows generally, as time passes, a steeper profile with respect to the gas profiles due to the inward drift of pebbles (increased $\Sigma$ in the inner part, decreased in the outer part of the disc).

The bottom panel of Fig. \ref{fig:surface_density} shows the scenario in which we allow planetesimal formation. The planetesimal surface density (red line) also presents spikes at the evaporation fronts, due to the re-condensation of gas forming a higher density of pebbles, which leads to the formation of planetesimals.
As observed in \citet{Lenz2019}, the planetesimal surface density profile is steeper than the initial dust and gas surface density. This happens in the case of not too high turbulence when the planetesimal formation is mostly hindered by the radial drift barrier, because the particles that are not converted into planetesimals in the outer part of the disc drift inwards and can still participate in the planetesimal formation in the inner part of the disc. Due to the formation of planetesimals, the pebble surface density is lower compared to the scenario without planetesimal formation.
This effect could also be important to explain the abundance difference of the binary stars HD106515. In that system, one star hosts a giant planet, while the other has no detected planet. In order to explain the peculiar oxygen abundance difference, the disc around the star that does not form a planet needs to form planetesimals efficiently in order to trap oxygen rich ices, relevant to explain the abundance differences \citep{Huehn_2023}.

\section{Atmospheric composition}
\label{app:atmo_comp}
   \begin{figure*}
   \centering
   \includegraphics[width = \textwidth, keepaspectratio]{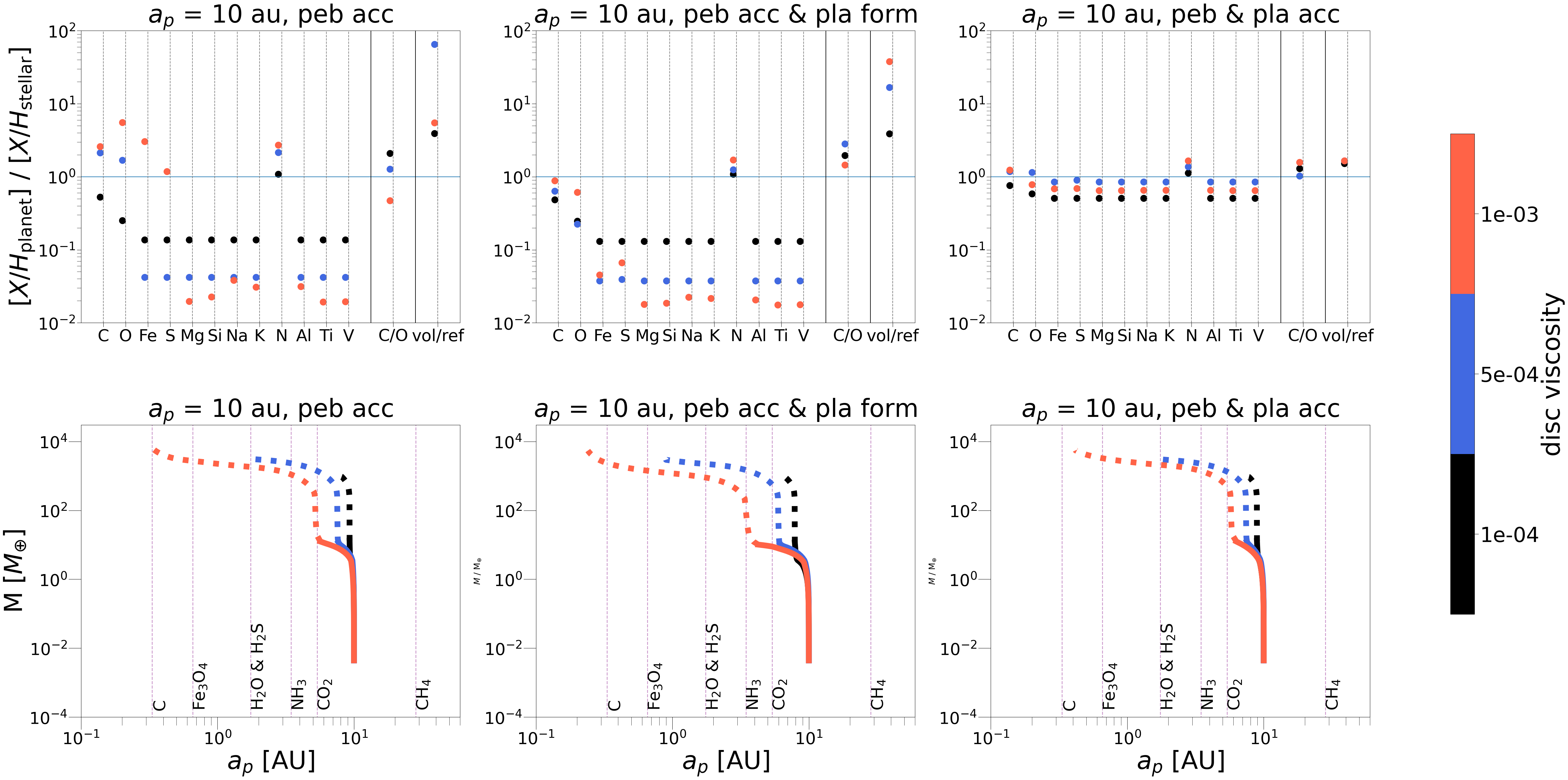}
   \caption{Same as Figure \ref{fig:atm_comp} but for planets starting at $10$ AU.}
        \label{fig:growth_tracks10}%
    \end{figure*}

   \begin{figure*}
   \centering
   \includegraphics[width = \textwidth, keepaspectratio]{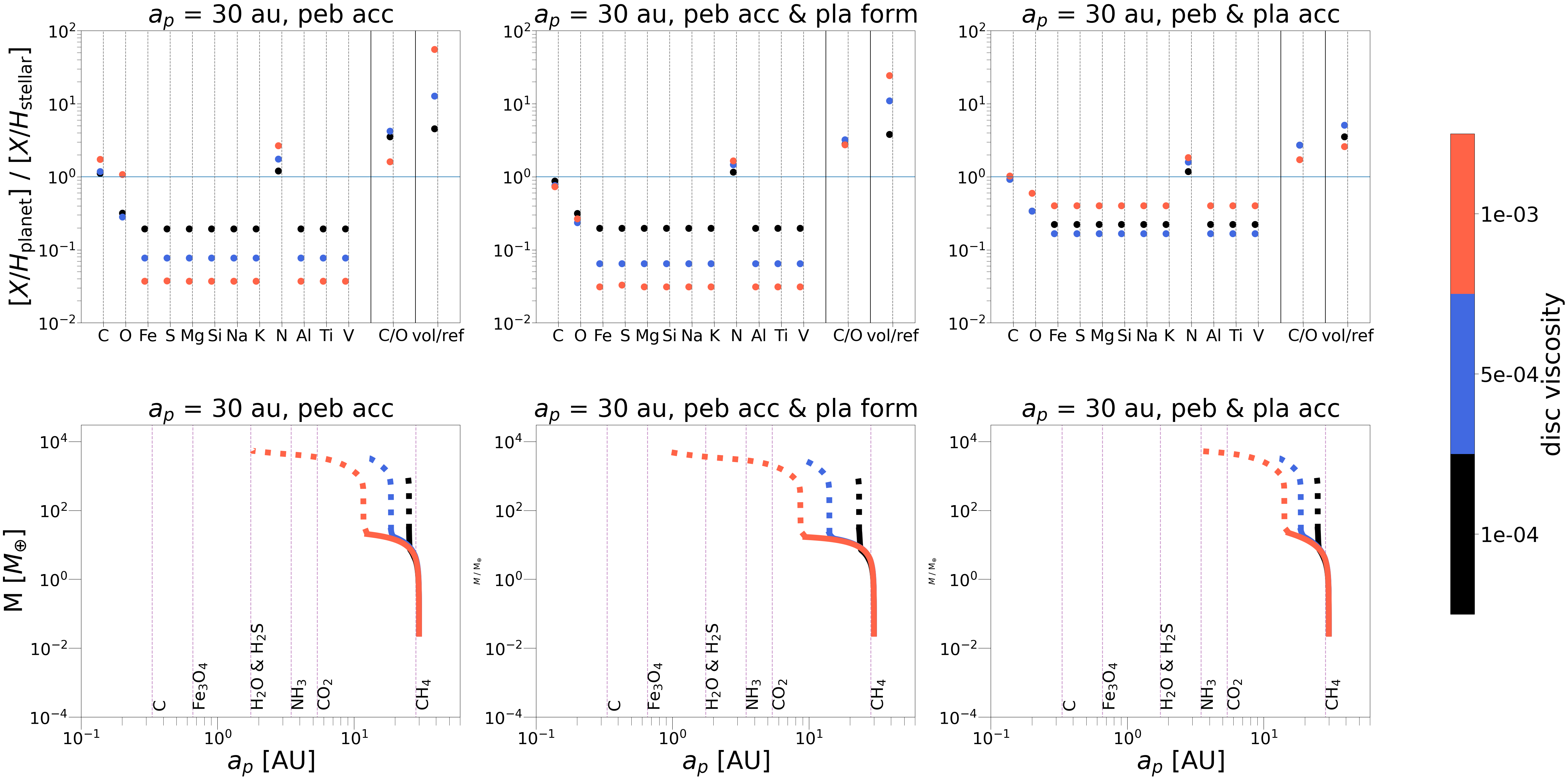}
   \caption{Same as Figure \ref{fig:atm_comp} but for planets starting at $30$ AU.}
        \label{fig:growth_tracks30}%
    \end{figure*}

Figures \ref{fig:growth_tracks10} and \ref{fig:growth_tracks30} show the normalized abundances of the chemical species, volatile to refractory ratio and C/O ratio for the $10$ and $30$ AU planets.
Compared to Fig. \ref{fig:atm_comp} that shows the same quantities for a $3$ AU planet, we observe that the elemental abundances are less enhanced in the pebble accretion only scenario because the planets start further out and therefore, by crossing less evaporation fronts, do not have the same chance to accrete enriched gas. The trend is clear also by observing the growth tracks in the second rows and comparing the high viscosity scenario (red lines), where the planets migrates much more towards the inner disc, to the low viscosity one (black lines), where the planets ends up staying further out.

In the case of planetesimal accretion, we observe for both, planets that start at 10 or 30 AU, a reduction in the volatile to refractory ratio as for the 3 AU planet, although the reduction for the  planet starting at 30 AU is smaller.

\section{Evolution of the atmospheric C/O ratio}
\label{app:growth_tracks}
   \begin{figure*}
   \centering
   \includegraphics[width = \textwidth, keepaspectratio]{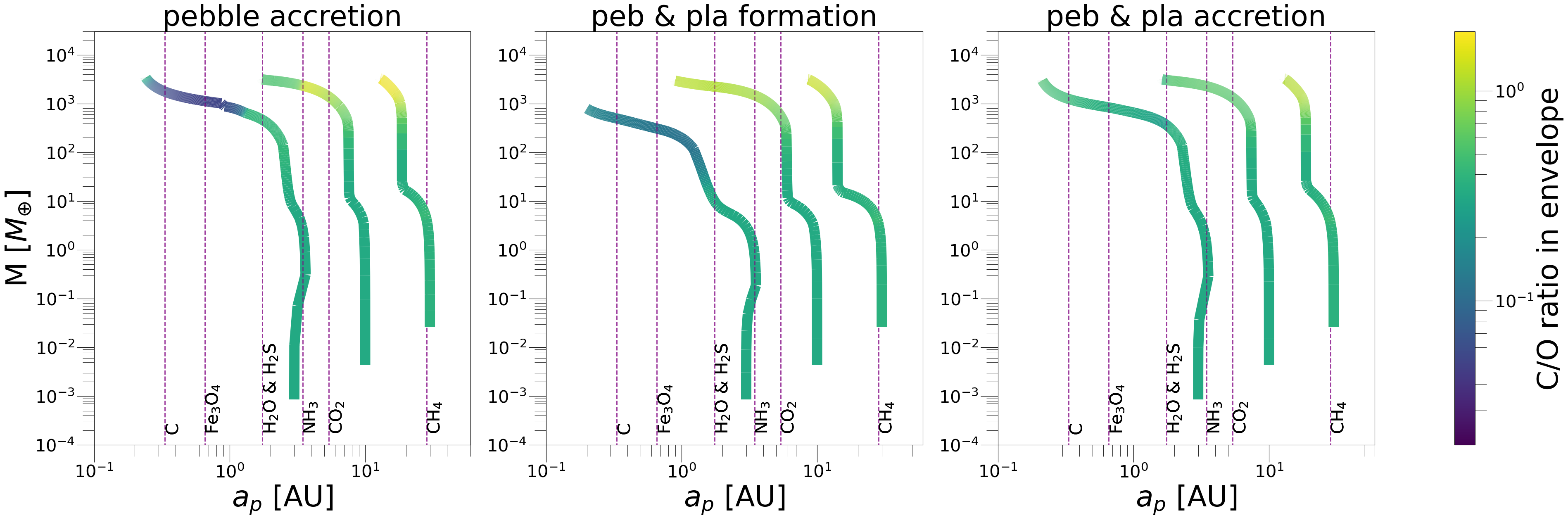}
   \caption{Growth tracks of the $3, 10, 30$ AU planets in the different scenarios (left to right) for a disc viscosity of $\alpha = 5 \cdot 10^{-4}$, where the colour coding represents the atmospheric C/O ratio.}
        \label{fig:growth_tracks}%
    \end{figure*}
    
Figure \ref{fig:growth_tracks} shows the growth tracks of the $3,10,30$ AU planets for the three different scenarios (left to right) with the C/O content in the envelope colour coded.

We observe that the atmospheric C/O ratio changes as the planet crosses the evaporation fronts. The most visible changes happen in the pebble accretion scenario, as the planet accretes highly enriched gas, while in the planetesimal formation scenario the changes are smaller, because the gas is less enriched in evaporated material.

Focusing on the crossing of the water evaporation front, we see that in the pebble accretion scenario there is a significant drop in the C/O ratio due to the accretion of water-rich vapour. In the planetesimal formation scenario this drop is less visible because the gas is less enriched, while in the planetesimal accretion scenario, the drop is sensitively smaller because, as the planet crosses the water evaporation front, it accretes water enriched vapour but it also accretes planetesimals from that location, that are instead carbon rich,  due to the large fraction of refractory carbon grains in our model. The final C/O content of the atmosphere is slightly different for the three scenarios, but depends on many parameters, thus making it difficult to distinguish between the formation scenarios via the atmospheric C/O ratio alone.

\section{Heavy element content origin}
\label{app:heavy}

\begin{figure*}
     \centering
     \begin{subfigure}[b]{0.49\textwidth}
         \centering
         \includegraphics[width=\textwidth]{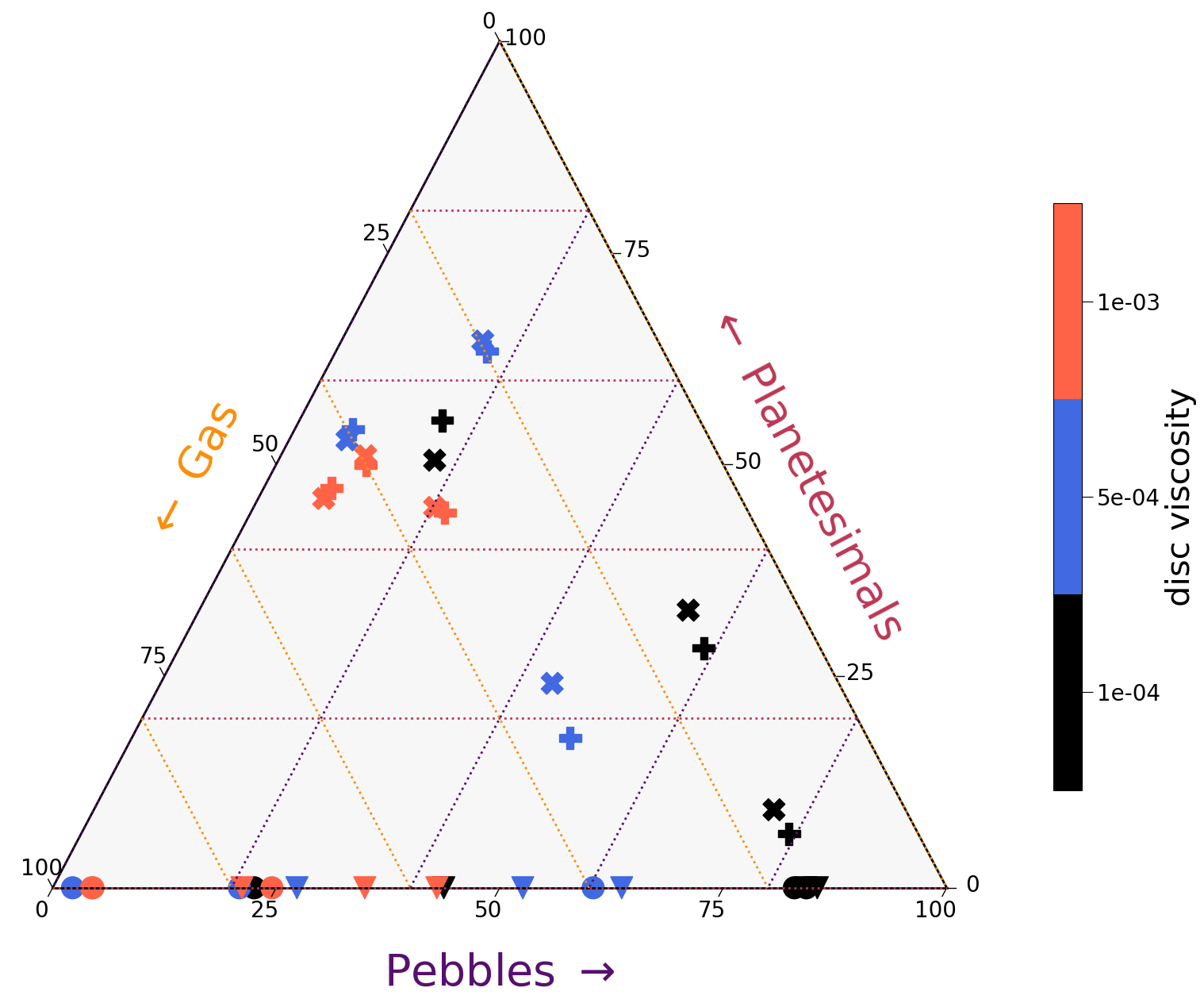}
         \caption{}
         \label{fig:ternary_visc}
     \end{subfigure}
     \hfill
     \begin{subfigure}[b]{0.49\textwidth}
         \centering
         \includegraphics[width=\textwidth]
         {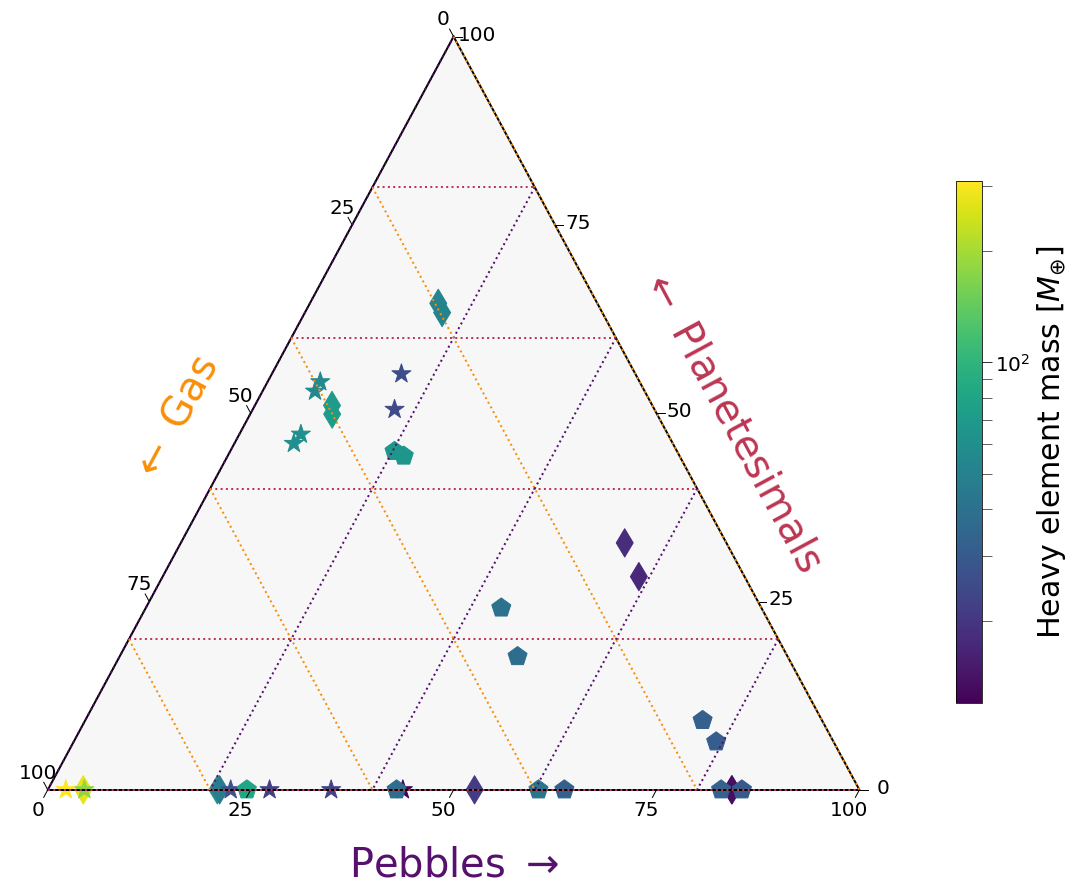}
         \caption{}
         \label{fig:ternary_heavy}
     \end{subfigure}
        \caption{Heavy element mass origin for the $3,10$ and $30$ AU planets. Panel \ref{fig:ternary_visc}: heavy element mass origin for some of the simulated planets, with the different disc viscosities colour coded. The different markers represent the different scenarios: pebble accretion (dots), planetesimal formation (triangles) and planetesimal accretion (plus = 50 km, crosses = 1 km planetesimals). Panel \ref{fig:ternary_heavy}: heavy element mass origin for some of the simulated planets, with total heavy element mass colour coded. The different markers represent the different initial positions of the planets: $3$ AU (stars), $10$ AU (diamonds), and $30$ AU (pentagons).
    }
        \label{fig:ternary}
\end{figure*}

The triangular plots in Fig. \ref{fig:ternary} shows the origin of the heavy element content of the $3, 10$ and $30$ AU planets in the three different scenarios for the different viscosities. Each of the three axes represents the percentage of heavy element mass provided by a certain phase (gas, pebbles, or planetesimals). 
In the left plot (Fig. \ref{fig:ternary_visc}), the different markers represent the three different scenarios: the dots for the pebble accretion only scenario, the triangles for the planetesimal formation scenario, and the plus and crosses for the planetesimal accretion scenario.
In the right plot (Fig. \ref{fig:ternary_heavy}), the different markers represent the three different initial positions of the planets: the stars for the $3$ AU planets, diamonds for the $10$ AU and pentagons for the $30$ AU, while the colour bar represents the heavy element content.

All the planets simulated in the first two scenarios lie on the bottom row because they do not accrete planetesimals. We observe that in the pebble accretion only scenario (dots) most of the heavy element mass of the planet is in gaseous form (cfr. the two $3$ AU planets in the bottom left corner with more than $90\%$ of heavy elements in the gas phase), due to the planets migrating in the very inner disc region that is heavily enriched in vapour. Planets forming in the outer disc region, on the other hand, seem to be more dominated by pebble accretion rather than vapour accretion. The reason is twofold: i) the pebble isolation mass is larger in the outer disc regions, resulting in larger cores and thus accreted pebbles and ii) as the outer disc is less enriched in vapour, due to the cold temperatures that allow many volatiles to be present as ices, the planets can accrete less heavy elements through the gas phase.

In the planetesimal formation case, instead, the pebble contribution to the heavy element content is more significant (between $25 \%$ and $75 \%$). This is caused by the fact that planetesimal formation takes away pebbles resulting in a smaller disc enrichment with vapour, while the core mass is similar to the pebble only scenario, resulting in larger fraction of pebbles within the total heavy element content.

Planets simulated in the planetesimal accretion scenario gain most of their heavy element mass from planetesimals, where more than $50 \%$ of the heavy mass can be due to planetesimal accretion. This is due to the fact that the heavy elements are locked into planetesimals and therefore cannot enrich the gas and be accreted in gaseous form, but are then dumped into the planet when planetesimals are accreted.

In the planetesimal accretion scenario, the $3$ AU planets are all concentrated in the same part of the diagram, regardless of the viscosity or the planetesimal radius. The $10$ AU planets show a smaller planetesimal and gas mass fraction for low viscosities and a higher one for higher viscosities, while the $30$ AU planets are the ones with the lowest planetesimal fraction. This is caused by the fact that the outer disc harbours a low planetesimal surface density, preventing an efficient accretion.

We also observe a trend both in the $10$ end $30$ AU planets: the final total heavy element mass increases with increasing viscosities. This is caused by the fact that the outer planets in the low viscosity environments only migrate very little and therefore stay in the outer disc. Consequently, they have only access to small amounts of planetesimals and additionally the disc is not enriched to large values with vapour, because the planets are exterior to the main evaporation fronts of water and CO$_2$.
\end{appendix}

\end{document}